# SAN/AFS: Developments in Storage Data Systems on Frascati Tokamak Upgrade

A. Bertocchi, G. Buceti, C. Centioli, F. Iannone, M. Panella, V. Vitale
Associazione Euratom/ENEA sulla Fusione,
C.R. ENEA Frascati – Rome, Italy

Abstract

In the last three years, the architecture of Frascati Tokamak Upgrade (FTU) experimental database has undergone meaningful modifications, data and codes have been moved from mainframe to UNIX platforms [1] and AFS (Andrew File System) has been adopted as distributed file system.

Further improvement we have added regards data storage system; the choice of SAN (Storage Area Network) over Fiber Channel, combined with power and flexibility of AFS, has made data management over FTU very reliable.

Performance tests have been done, showing better transfer rate than previous system, based on JBOD modules with SCSI connection.

## 1 INTRODUCTION

During last year FTU, Frascati Tokamak Upgrade, a compact high magnetic field tokamak, has reached the complete working configuration with three auxiliary heating systems to investigate radio frequency power deposition. It operates producing 20-25 shots per day, each shot being 1.5 s. long and producing presently about 30-40 MB of data for 1400 channels. This paper will discuss FTU data distributed architecture and storage systems solution.

## 2 FTU STORAGE SYSTEMS

The IBM-mainframe to UNIX migration of data and analysis software in FTU was been done trying several storage systems and server hardware platform. The main item of new operating environment was to adopt a scalable file sharing in multi-platform data processing.

Initially the 150 Gbytes of FTU data shots was stored on the RAID Array 450 of DIGITAL UNIX using RAID 5 raidset to have a proper data reliability and a good transfer rate in read data. Unfortunately, this solution was not performance under shared file system (AFS 3.5/6) because of the insufficient memory cache of the raid controller (32 Mbytes) and a very poor transfer rate in write data. We substituted the RAID array with a SCSI JBOD (*Just a Bunch of Disks*) type IBM 36 Gbytes 10000 rpm and Seagate 50 Gbytes 7000 rpm, both SCSI Ultra Wide. No raidset level was configured for that storage system to require a good transfer rate, in spite of the fact that losing redundancy means low data reliability. Subsequently we evaluated a new storage product considered as one of the hottest topics in IT: Storage Area Network (SAN) over Fibre Channel (FC).

SAN, a high-speed, high-performance network, permits multi-platform servers with heterogeneous operating system to access to multi-vendor storage devices as equals; FC allows to attach up to 126 devices over a range of 10 kilometres (in fibre optic) with ~100 MB/sec data transfer rate, exceeding SCSI limitations.

### 2.1 RAID systems

We replaced the provisional JBOD storage system with a Dot-Hill's SANnet storage system that offers a family of advanced disk storage products for open systems servers.

SANnet 4200 is an external hardware RAID storage system that can be used with stand-alone servers, server clusters, or as part of a SAN. It includes:

1. Two FC host channels providing throughputs ranging from 100 MB/s in single host server to 200 MB/s for clustered host configuration and five 80 MB/s Ultra SCSI (LVD) disk drive channels
2. Four FC connections that support single or multiple servers simultaneously and are compatible with many server platforms including Unix, Linux and Windows
3. A capacity from 2 to 10 disk drives standard, expansible up to 50 disk drives with expansion system
4. A storage capacity that ranges from 18 GB to 9 TB and support disk drive of: 9.2, 18.4, 36.7, or 73 GB with rotational rate up to 10000 RPM and 180 GB with 7200 RPM

In addition it allows:
1. High reliability because of a redundant hot-swap removable components: disk drivers, RAID controller, battery backups, event reporting cards, power supply and cooling fans
2. Single or dual redundant (active-active) RAID controllers with RAID levels: 0, 1, 0+1, 3, 5, 10, 30 and 50; RAID controllers can be equipped up to 1024 MB of cache memory per system (512 MB per RAID controller in active-active redundant FC networks) with automatic fail-over and fail-back.

A SANnet 4200 can be configured by first creating Logical Drives (LD), then partitions and lastly mapping host LUNs (Logical Unit Number).

SANnet 4200 is a high-reliable system that supports three levels of redundancy: 1) spare drive, 2) active-active redundant controller, 3) multi-servers operation.

At first level, hot swapping is supported through automatic disconnection of a failed drive and detection of a reserved Local (to serve a specified LD) or Global (to serve multiple LD) Spare Disk, followed by automatic background rebuilding of data.

At the second level, the active-active redundant controller means is the back up mechanism in case of failure of the primary RAID controller and it became necessary for the redundant controller to take over all servers servicing control.

The last level can be achieved in a switched topology: a redundancy of servers means that a RAID controller - in point-to-point otherwise loop - will attempt to login a public loop and the partitioned LD can be simultaneously mounted by many servers.

### 2.2 FTU Storage architecture

Presently, we have collected about 200 GB of experimental data, including 140 GB of FTU data (raw and processed) and 60 GB of users data. To storage them we have installed two RAIDs SANnet 4200 in *single controller* configuration, each one with ten 73 GB (UW SCSI LVD, 10000 RPM) Seagate disks for a total capacity of 1.46 TB. Each RAID controller has 256 MB of cache memory.

We adopted a *point-to-point otherwise loop* topology, with four SUN ULTRA 10 Servers (CPU UltraSPARC-Iii, 440 Mhz, 256 MB of RAM and SunOS 5.8) with FC Host Bus Adapter Emulex LP 8000 in DE9 layout (copper duplex cabling with DB9 connectors).

We configured the SANnet systems in RAID 5 level with one Global Spare Disk for each RAID. It has been created two Logical Drives with capacity of 280 GB and 209 GB respectively, each one mapped on two different host channels. In this way, each server mounts a disk partition of about 200 GB and it is possible to achieve a full throughput of 200 MB/s for each RAID controller. An Automatic Tape Library – ATL – (15 AIT Cartridge Treefrog of Spectralogic) with differential SCSI adapter is used for cold data back up. The ATL has a maximum native capacity of 375 GB and a maximum compressed capacity of 975 GB. Presently SCSI adapter directly connects the ATL to a server, but it is possible to move to FC link and to connect to FC-switch. In this way, all servers can back-up their data through SAN without to interfering the traffic on the LAN.

Fig.1 shows a schematic block diagram of FTU storage systems.

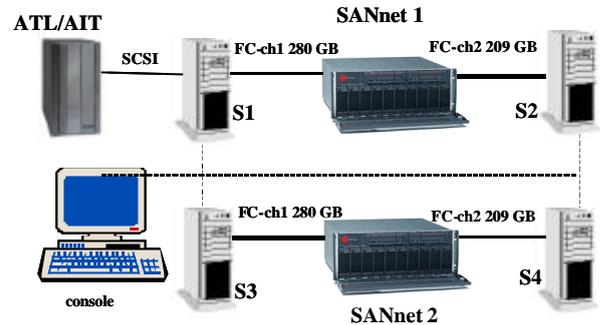

Figure 1: FTU Storage system

To show how the reliability has a cost, it should be noted that at the first level of redundancy, between Global Spare Disk and RAID 5 level, the native storage capacity of 730 GB of each RAID decreases to 489 GB with a capacity lose of 33 %.

Furthermore, to upgrade to the second level of redundancy, where it is adopted a dual RAID controller configuration, it is needed to add a secondary controller with 512 MB of cache memory and to expand to 512 MB the cache memory of primary controller, with a cost that is about 60 % of RAID system in single controller configuration.

### 2.3 Bechmarks

To confirm the right choice of new architecture, we have carried out transfer rate tests on two different systems. Benchmarks consisted of moving data volumes including 100 shots (130124 files and 6057 folders, about 1.52 GB of data) from one directory to another and vice versa.

Table 1 reports test results; the RAID system based on SANnet 4200 has a better transfer rate (between 26% and 31%) than previous one, based on JBOD modules with SCSI connection.

Table 1: Transfer rate benchmarks

| Storage System | Move A -> B (mm:ss) | Move B -> A (mm:ss) |
|---|---|---|
| JBOD (#2 36 GB, 10000 rpm, UW SCSI IBM; #2 50 GB, 7200 rpm, UW SCSI Barracuda Seagate) | 35' 55" | 35' 10" |
| SANnet (#20 73 GB, 10000 rpm, UW SCSI3 Seagate) | 26' 39" | 24' 24" |

## 3 DISTRIBUITED FILE SYSTEMS

The adoption of AFS (*Andrew File System*) as client/server architecture allows to access files and directories on geographically distributed machines like a unique virtual machine, under the */afs* file system. An AFS client can approach FTU experimental data simply connecting under */afs*, where a single channel/file is addressed as:

*/afs/fus/project/ftudati/das/hundred/number/family name/channel name*

regardless platform, OS and real physical data location. Fig.2 shows the AFS *fusione.it* cell structure.

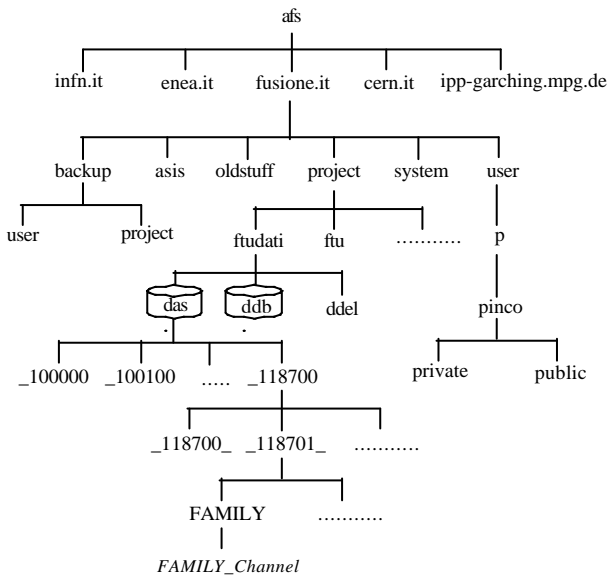

Figure 2: AFS *fusione.it* cell structure

Because AFS automatically replicates information across multiple servers within the enterprise to eliminate single points of failure and uses client-side caching which helps to reduce server and network loads, AFS provides uninterrupted access to data during routine maintenance, backup, or temporary system failures. Combined with powerful SAN features, this solution supplies full data availability.

## 4 CONCLUSIONS

From the lack of on line space and the need of tape robot in the mainframe environment, the fast change in the storage technology has moved the FTU archive to be completely on line. The flexibility of AFS has given us the chance to test several storage architectures until the adoption of SAN. The modularity of this solution allows us to face any predictable future space need for the FTU experiment.